\documentclass[11pt,oneside,letterpaper]{article}
\usepackage{amssymb}
\usepackage{amsmath}
\usepackage{graphicx}
\usepackage{setspace}
\usepackage{fancyhdr}
\usepackage{ifpdf}
\usepackage{graphicx}
\usepackage{comment}

\def\half{{\textstyle{1\over2}}}

\def\x{{AdS}}

 \def\p{\partial}

\def\ip{{\cal I}^+}

\newcommand{\bea}{\begin{eqnarray}}
\newcommand{\eea}{\end{eqnarray}}
\newcommand{\be}{\begin{equation}}
\newcommand{\ee}{\end{equation}}

\newcommand{\bi}{\begin{itemize}}
\newcommand{\ei}{\end{itemize}}

\newcommand{\qed}{\nobreak \ifvmode \relax \else
      \ifdim\lastskip<1.5em \hskip-\lastskip
      \hskip1.5em plus0em minus0.5em \fi \nobreak
      \vrule height0.75em width0.5em depth0.25em\fi}

\numberwithin{equation}{section}
\allowdisplaybreaks  

\begin{document}
\begin{titlepage}
\centerline{}
\vskip 2in
\begin{center}
{\LARGE{\textsc{ Future Boundary Conditions in \\ De Sitter Space}}}
\vskip 0.5in
{Dionysios Anninos, Gim Seng Ng and Andrew Strominger}
\vskip 0.3in
{\it Center for the Fundamental Laws of Nature,
Harvard University\\
Cambridge, MA, 02138

}
\end{center}
\vskip 0.5in
\begin{abstract}
We consider asymptotically future de Sitter spacetimes endowed with an eternal observatory. In the conventional descriptions, the conformal metric at the future boundary $\ip$ is deformed by the flux of gravitational radiation. We however impose an unconventional future ``Dirichlet'' boundary condition requiring that the conformal metric is flat everywhere except at the conformal point where the observatory arrives at $\ip$. This boundary condition violates conventional causality, but we argue the causality violations cannot be detected by any experiment in the observatory. We show that the bulk-to-bulk two-point functions obeying this future boundary condition are not realizable as operator correlation functions in any de Sitter invariant vacuum, but they do agree with those obtained by double analytic continuation from anti-de Sitter space.
\end{abstract}

\end{titlepage}

\setcounter{page}{1}
\pagenumbering{arabic}

\tableofcontents

\section{Introduction and summary}

The discovery of a nonzero cosmological constant suggests that our universe asymptotes to a de Sitter (dS) spacetime  in the infinite future ($\ip$). Interestingly, the proper characterization of dynamics in such asymptotically future dS spacetimes - the analog of the S-matrix for asymptotically Minkowski spacetimes or boundary correlators for asymptotically anti-de Sitter (AdS) spacetimes - remains an open problem. In this paper we consider such dS spacetimes and  endow them with with an eternally-funded observatory whose fattened worldline is denoted $\cal W_O$.  We use the word observatory rather than observer to emphasize that we are considering an object of finite  extent in which arbitrary experiments can be performed and indefinitely repeated. We explore herein the premise that these experiments comprise the basic dS observables, and that correlators on the cylindrical boundary of $\cal W_O$ play a role in dS dynamics in some respects akin to the role played by correlators on the asymptotic cylindrical boundary of AdS in AdS dynamics.\footnote{This approach to dS has been advocated elsewhere including  \cite{Gibbons:1977mu,Alishahiha:2004md,Banks:2003cg,Banks:2010tk,Goheer:2002vf,Castro:2011xb}. One objection to it is that an observatory which can record and store all such information must have an infinite number of microstates, while de Sitter space itself might have only finitely many \cite{Banks:2002wr}. }

Of course dS has an asymptotic  future boundary - $\ip$ - which in some ways resembles the asymptotic spatial boundary of AdS, with the role of space and time reversed. However there are several key differences between these boundaries. One is that correlators on $\ip$ of dS cannot be measured by any physical experiment because all points on $\ip$ are causally disconnected.\footnote{This of course would not apply to an early-time approximately dS inflationary phase of our universe--see below.} For this reason they are sometimes referred to as ``metaobservables"  \cite{Witten:2001kn}.  A second difference is that Dirichlet-type boundary conditions on the metric and other fields can be imposed at the asymptotic boundary of AdS, ensuring that no radiation passes out of the boundary of the spacetime and the energy is conserved.  Imposing such Dirichlet-type boundary conditions at dS $\ip$  would violate causality and lead to inconsistencies with the usual type of dS initial value formulation on complete spacelike slices.  In general we expect radiation can pass through $\ip$ and the charges are not conserved. The asymptotic structure of dS was recently analyzed in \cite{Anninos:2010zf}, where it was shown that the asymptotic symmetry group (ASG) is all the diffeomorphisms tangent to  $\ip$, and the associated charges obey  a conservation law relating their variation to the radiation flux through $\ip$. This is in marked contrast to AdS where the ASG is the finite-dimensional (for $D >3$) conformal group. The dS case resembles more the case of Minkowski $\ip$ whose ASG is the infinite-dimensional BMS group \cite{BM62,S62a}.

In this paper however, we question the notion that one should think about dS dynamics in terms of imposing initial data (or a quantum state) on a complete spacelike slice and then evolving it to the future. While mathematically well-defined, this is highly unphysical, since such slices necessarily contain causally disconnected regions. Hence the resulting spacetime does not correspond to anything which could be physically measured.

Here we propose that initial data should instead be imposed on the boundary $\p{\cal W_O}$ of the fattened observatory worldline. In the gravity sector, which is esentially all we consider here, this means specifying the intrinsic metric and extrinsic curvature, subject of course to the constraint equations, on the cylindrical timelike hypersurface $\p{\cal W_O}$. Physically this means we are characterizing the spacetime by what passes in and out of the observatory walls -- clearly measurable data. To determine the bulk geometry, we must evolve radially outward rather than forward in time.

The radial evolution of this initial data on $\p{\cal W_O}$ may fully determine the geometry within the ${\cal W_O}$ causal diamond but not  on $\ip$. We propose to fix the $\ip$ geometry by imposing Dirichlet boundary conditions on the conformal metric everywhere except at the conformal point $\cal P_O$ where ${\cal W_O}$ reaches $\ip$. This condition together with the $\p{\cal W_O}$ initial data  remaining data on $\ip$ - which turns out to be the  conformal traceless part of the extrinsic curvature - plausibly determines the entire spacetime.\footnote{In the case of vanishing cosmological constant, the problem of data on a worldtube and initial null slice was formulated in a very similar fashion in \cite{TW66}.}
The main - if simple - point of this paper is that, while such a boundary condition  is manifestly acausal, the causality violations are apparently unobservable, $i.e.$  they cannot be detected by any physical experiment in the observatory.  We show explicitly at the linearized level that the future boundary condition  imposes no acausal restriction on the initial data on $\p{\cal W_O}$, and that these conditions together determine the full dS geometry. We expect these results to extend beyond the linearized level to a finite neighborhood of the vacuum dS geometry.

To understand why this is possible, consider a gravity wave produced at the observatory which passes through the future ${\cal W_O}$ horizon and reaches $\ip$. Dirichlet boundary conditions will acausally reflect it backward in time, but the reflected wave remains outside the ${\cal W_O}$ causal diamond. Another way of thinking of this is that the boundary condition acausally places ``de Sitter demons" outside the ${\cal W_O}$ causal diamond. Every time a wave comes out of the observatory, a de Sitter demon sends a finely-tuned wave to $\ip$ which interferes destructively with the observatory wave so as to maintain the Dirichlet boundary condition on $\ip$.

Going beyond pure gravity, we expect this type of boundary condition makes sense in theories with no massive particles or black holes which are absolutely stable. Everything must ultimately decay to massless particles. If a localized stable  object reaches $\ip$ the future boundary condition cannot be maintained by the mechanism discussed here - although there may be a generalization.

The theory of inflation proposes  that our universe had a long  era in which the geometry was very close to dS with a large cosmological constant.  The considerations of this paper do not directly apply to this era. We are metaobservers for this early dS phase: we can see events which would have been forever causally disconnected had there been no exit from inflation. Indeed the CMB and its fluctuations can be  approximately thought of as the correlation functions on the would-be $\ip$ of the early dS phase \cite{Maldacena:2002vr}. There is also no horizon or Bekenstein-Hawking entropy associated to the early phase once the exit from inflation into the present phase is taken into account. Clearly there are qualitative differences between an early-time and future asymptotic dS phases. It would be interesting to understand how the description of one goes to the other as the lifetime of the phase becomes infinite.

The considerations of this paper are purely classical and we do not attempt to define a quantum theory consistent with the future boundary conditions.  Nevertheless our observations may have implications for attempts to construct a holographic dual for dS quantum gravity.  This is of course a wide open problem. It is not even clear where the best home for the dual is: $\ip$, the horizon and $\p{\cal W_O}$ are among the possibilities. One might expect the ASG for dS to be the symmetry group of the dual theory.  Taken at face value, the result of \cite{Anninos:2010zf} that the ASG is all diffeomorphisms of $\ip$ suggests that  the dual should itself be a theory of gravity in one lower dimensions. This large ASG came from the absence of $\ip$ boundary conditions in the usual approach. If we apply Dirichlet boundary conditions at $\ip$, as in the present paper, the structure becomes very similar to that of AdS.  Indeed the dS two-point functions with these boundary conditions are precisely the analytic continuations (in the cosmological constant) of the AdS two-point correlation functions, and transform under the Euclidean conformal group $SO(D,1)$. This suggests that the  holographic dual is a conformal field theory without gravity, as envisioned in the dS/CFT correspondence \cite{Strominger:2001pn, Strominger:2001gp}. Hence the boundary conditions proposed herein brings the structure of dS much closer to that of AdS, and hopefully will be useful in adapting insights from AdS holography to the dS context.

Our results resonate with a recent paper  \cite{Maldacena:2011mk} considering future boundary conditions for conformal gravity in dS.  It was shown that they can be chosen to classically reduce dS conformal gravity to dS Einstein gravity.  This reduction however requires future boundary conditions everywhere on $\ip$ and might be ruined by the exclusion of the point $\cal P_O$. Nevertheless our observations may be relevant to a better understanding of the relation between conformal and Einstein gravity. Our picture may also bear some relation to Schrodinger's $Z_2$ antipodal identification of dS \cite{es,Parikh:2002py,Sanchez:1987sa} or black hole final state boundary conditions \cite{Horowitz:2003he} and is in the spirit of black hole complementarity \cite{Susskind:1993if}.

This paper is organized as follows. In section 2 we consider as a warmup the case of light scalars  in dS$_3$. Below a critical value of the mass these modes, like 4D gravitons, do not oscillate at $\ip$ and can have a slow or fast exponential falloff. We show that, for any mode sourced in the southern causal diamond, demons located in the causally complementary northern diamond can excite a northern mode which will interfere with the southern mode in such a way that the total mode has only the fast-falling component near $\ip$. The phase of the northern demon mode depends on the mass, angular momentum and frequency of the southern mode.
In section 3 we show that northern demons in dS$_4$ can similarly destroy the slow falling components of
gravity waves produced in the southern diamond. This means that the conformal metric on $\ip$ retains its round shape, although the conformal extrinsic curvature is altered by the wave. In section 4 we consider two-point functions in our setup, returning for simplicity to the case of light dS$_3$ scalars. We show that a unique symmetric two-point function is determined by demanding dS invariance, fast falloff at $\ip$ and Hadamard form of the coincident-point singularity. We further show that this two-point function can $not$ arise as the Wightman function of  two scalar fields in any quantum state defined on complete spacelike slices in dS,
but $does$ result from double analytic continuation of the standard AdS scalar two-point function.
The appendix contains a construction of the dS$_4$ graviton modes in global coordinates, complementing the static patch analysis of section 3.

\section{Warmup: light scalars in dS$_3$}

In this section we study light scalars with masses less than the critical value $\mu^2=\Lambda/3$.  The $\mu^2>\Lambda/3$ case was studied with similar conventions in \cite{Bousso:2001mw}.  The asymptotic behavior of such light scalars resembles 4D gravitons in that they exponentially decay rather than oscillate near $\ip$. We use the metric in static patch coordinates:
\be
\frac{ds^2}{\ell^2} = -(1-r^2)dt^2 + \frac{dr^2}{(1-r^2)} + r^2 d\varphi^2~.
\ee
where $\ell^2 = 3/  \Lambda$. The southern causal diamond associated to an observatory at the south pole is described by $r\in [0,1]$. The northern causal diamond which will be populated by demons is described by a second copy also with  $r\in [0,1]$. We take time to  run forward in the southern diamond and backwards in the northern diamond so that $\p_t$ is the globally defined Killing vector. The future and past diamonds, containing $\mathcal{I}^+$ and $\mathcal{I}^-$ respectively, are described by $r \in [1,\infty]$. In the future diamond, the spacelike $t$-coordinate runs from  north to south, whereas in the past diamond it runs south to north.

\subsection{Northern and southern modes}

The scalar field modes may be labeled by the angular momentum $j$ in the $\phi$ direction and the frequency $\omega$ in time. The  general solution of the scalar wave equation for mass $0 < \mu^2\ell^2 < 1$ in the southern patch is then given by:
\be
\phi^S (t,r,\varphi) = \sum_{j \in \mathbb{Z},\omega>0} \left( \kappa_{j\omega} \phi^S_{\omega j} (t,r,\varphi) + \kappa^*_{\omega j} \phi^{S*}_{\omega j}(t,r,\varphi)  \right)~,
\ee
where the static patch modes smooth at the origin\footnote{Other types of behavior at the origin might be considered depending on the nature of the observatory stationed there.} are:
\be
\phi^{S}_{\omega j} = e^{-i\omega t + i j \varphi}~r^{|j|} (1-r^2)^{i\omega/2} F(a,b;c;r^2)~,
\ee
and the arguments of the hypergeometric function $F(a,b;c;r^2)$ are:
\be
a \equiv \frac{1}{2}\left( |j| + i \omega + h_+ \right)~,\quad b \equiv \frac{1}{2}\left( |j| + i \omega + h_- \right)~,\quad
c \equiv 1 + |j|~.
\ee
with
\be  h_\pm \equiv 1 \pm \sqrt{1-\mu^2\ell^2} .\ee   Note that $h_\pm$ are both real and positive in the mass range under consideration. There is a similar expansion for the northern modes, since the northern patch is described by an identical coordinate system with time running backwards.

\subsubsection*{Near the cosmological horizon}

We now study the behavior of the static patch modes near the cosmological horizon $r=1$. In Kruskal coordinates:
\be\label{kruskS}
r = \frac{1+UV}{1-UV}~, \quad t = \frac{1}{2} \log \left( - \frac{U}{V} \right)~,
\ee
the southern diamond is the region $U>0$, $V<0$ and the future (past) horizon is at $V=0$ ($U=0$). Using the hypergeometric identity:
\begin{multline}\label{hypident}
F\left( a,b;c;z \right) =  \frac{\Gamma(c)\Gamma(c-a-b)}{\Gamma(c-a)\Gamma(c-b)} F\left( a,b;1+a+b-c;1-z \right)  \\
+   \frac{\Gamma(c)\Gamma(a+b-c)}{\Gamma(a)\Gamma(b)} (1-z)^{c-a-b} F\left( c-a,c-b;c-a-b+1;1-z \right)
~,
\end{multline}
the near horizon behavior is
\be
\phi^S_{\omega j}\sim e^{ i j \varphi} \left[ \alpha_{\omega j} (-V)^{i\omega} + \alpha_{-\omega j} U^{-i\omega} \right]~,
\ee
with:
\be
\alpha_{\omega j} \equiv \frac{\Gamma(1+|j|)\Gamma(-i \omega)2^{i\omega}}{\Gamma\left(\frac{1}{2}(|j|-i\omega+h_+)\right)\Gamma\left(\frac{1}{2}(|j|-i\omega+h_-)\right)}~.
\ee

\subsection{Future and past modes}

In the future diamond, we can build $\phi^{out \pm}_{j\omega}$ modes that behave as $\sim r^{-h_{\pm}}$ near $\mathcal{I}^+$. Explicitly we find for the fast-falling $out+$ modes:
\be
\phi^{out+}_{\omega j} =  e^{-i\omega t + i j \varphi}~r^{-h_+}\left(1-\frac{1}{r^2}\right)^{i\omega/2} F(a,1-a^*_-;h_+;\frac{1}{r^2})~,
\ee
where $a^*_-$ is given by taking the expression for $a^*$ and replacing $h_+$ with $h_-$. For the slow-falling $out-$ modes we find:
\be
\phi^{out-}_{\omega j} =e^{-i\omega t + i j \varphi}~r^{-h_-}\left(1-\frac{1}{r^2}\right)^{-i\omega/2} F(a^*_-,1-a;h_-;\frac{1}{r^2})~.
\ee

\subsubsection*{Near the cosmological horizon}

Once again, we can expand the above expressions for the $out$ modes near the cosmological horizon. The Kruskal coordinates are now given by:
\be\label{kruskout}
r = \frac{1+ U V}{1 - U V}~, \quad t = \frac{1}{2}\log\left( \frac{U}{V} \right)~,
\ee
where $U>0$ and $V>0$ in the future diamond. The $out+$ modes behave as:
\be
\phi^{out+}_{\omega j} \sim e^{i j \varphi} \left[\beta_{\omega j} V^{i\omega} + \beta_{-\omega j}U^{-i\omega} \right]~,
\ee
with:
\be
\beta_{\omega j} = \frac{\Gamma(h_+)\Gamma(-i\omega)2^{i\omega}}{
\Gamma\left( \frac{1}{2}(h_+ -|j|-i\omega) \right) \Gamma\left( \frac{1}{2}( h_+ + |j| - i\omega ) \right) }~.
\ee
Similarly, the expansion of the $out-$ modes near the cosmological horizon is as above, but with $\beta_{\omega j}$ replaced by:
\be
\gamma_{\omega j} = \frac{\Gamma(h_-)\Gamma(-i\omega)2^{i\omega}}{
\Gamma\left( \frac{1}{2}(h_- + |j|-i\omega) \right) \Gamma\left( \frac{1}{2}( h_- - |j| - i\omega ) \right) }~.
\ee

\subsection{Matching the flux}
When we send out a wave from the northern or southern patch, it will generically contain both fast and slow-falling out modes.
Matching the flux across the future horizon determines the Bogoliubov transformation relating the northern and southern  modes to the $\pm out$ modes:\footnote{For heavy modes, with $\mu\ell > 1$, this can be found in \cite{Bousso:2001mw}.}
\begin{eqnarray}
\phi^S_{\omega j} &=& A^{11}_{\omega j} \phi^{out-}_{\omega j} + A^{12}_{\omega j} \phi^{out +}_{\omega j}~, \\
\phi^N_{\omega j} &=& A^{21}_{\omega j} \phi^{out-}_{\omega j} + A^{22}_{\omega j} \phi^{out +}_{\omega j}~.
\end{eqnarray}
with:
\begin{eqnarray}
N_{\omega j} A^{11}_{\omega j} &=&  \alpha_{-\omega j}\beta_{\omega j} ~,\\
N_{\omega j} A^{12}_{\omega j} &=&  -\alpha_{-\omega j}\gamma_{\omega j}~,\\
N_{\omega j} A^{21}_{\omega j} &=& -\alpha_{\omega j}\beta_{-\omega j} ~,\\
N_{\omega j} A^{22}_{\omega j} &=&  \alpha_{\omega j}\gamma_{-\omega j}~.
\end{eqnarray} where $N_{\omega j} \equiv \beta_{\omega j} \gamma_{-\omega j} - \gamma_{\omega j} \beta_{-\omega j} = i \sqrt{1-\mu^2\ell^2}/\omega$.

\subsection{Demonic interference}

Now we would like to demonstrate that  the slow-falling piece of any southern mode at $\mathcal{I}^+$ by a wave produced by a de Sitter demon in  the causally disconnected northern diamond. More precisely, if the observatory excites  a normalized southern mode $\phi^S_{\omega j}$, the coefficient of the $\phi^{out-}_{\omega j}$ mode at $\mathcal{I}^+$ will be $A^{11}_{\omega j}$. So the  demon must excite a northern mode with Fourier coefficient $- A^{11}_{\omega j} / A^{21}_{\omega j}$, to cancel the slow falling component of the incoming southern mode. Then, the coefficient of the fast falling mode $out+$ becomes:
\be
A^{12}_{\omega j} - \frac{A^{11}_{\omega j}}{A^{21}_{\omega j}} A^{22}_{\omega j}   = \frac{\Gamma(1+|j|)\Gamma\left(\frac{1}{2}(h_+ -|j|+i\omega)  \right)}{\Gamma(h_+)\Gamma\left( \frac{1}{2}(h_- + |j| + i\omega) \right)}~.
\ee
Alternatively, we can express the above as:
\be
A^{12}_{\omega j} - \frac{A^{11}_{\omega j}}{A^{21}_{\omega j}} A^{22}_{\omega j}   = A^{12}_{\omega j} \left( 1-e^{i \lambda_{\omega j}}\right)~,
\ee
where:
\be
\tan{\left(\lambda_{\omega j}/2\right)} \equiv \frac{\sin{\left(\pi \sqrt{1-\mu^2 \ell^2}\right)} \sinh{\left( \pi \omega\right)}}
{\left(-1\right)^j+\cos{\left(\pi \sqrt{1-\mu^2 \ell^2}\right)} \cosh{\left( \pi \omega\right)}}~.
\ee
Thus, the full mode with no slow-falling behavior near $\mathcal{I}^+$ is given by:
\be\label{good}
\phi^S_{\omega j} -  \frac{A^{11}_{j\omega}}{ A^{21}_{j\omega} } \phi^N_{\omega j}~,
\ee
where:
\be
-  \frac{A^{11}_{j\omega}}{ A^{21}_{j\omega} } =  \frac{\Gamma(\frac{1}{2}(h_- + j - i\omega))\Gamma(\frac{1}{2}(h_+ - j + i\omega ))}{\Gamma(\frac{1}{2}(h_- + j + i\omega))\Gamma(\frac{1}{2}(h_+ - j - i\omega ))} = e^{-i\delta_{\omega j}}~,
\ee
and $\tan(\delta_{\omega j}/2) \equiv \tan\left( \frac{\pi}{2} \left( j - \sqrt{1-\mu^2\ell^2} \right)\right)\tanh(\pi \omega/2)$.

\section{Linearized gravity in  dS$_4$}

We now consider the problem of linearized gravity in the static patch of dS$_4$, following the work of \cite{Kodama:2003kk}. The 4D dS  metric is :
\be
\frac{ds^2}{\ell^2} = - (1-r^2) dt^2 + \frac{dr^2}{(1-r^2)} + r^2 d\Omega^2_2~,
\ee
The linearized gravitational excitations can be parametrized by a transverse vector spherical harmonic and a scalar spherical harmonic. Together, these constitute two degrees of freedom. There is no transverse-traceless tensorial spherical harmonic for a two-sphere. Since the computation is essentially identical for both types of harmonics, we only consider the vector harmonics in what follows.

\subsection{Vector excitations}

We can express \cite{Kodama:2003kk} the vectorial perturbations in terms of a transverse vectorial spherical harmonic $\mathcal{V}_{i}$:
\begin{eqnarray}
\delta g_{ij} &=& 2 r^2 H_T (r,t) \mathcal{V}_{ij}~, \quad i,j \in \{ \theta, \phi \}~, \\
\delta g_{ai} &=&  r f_a \mathcal{V}_i~, \quad a \in \{ t,r \}~.
\end{eqnarray}
with all other components of $\delta g_{\mu\nu}$ vanishing. We have further defined:
\be
\mathcal{V}_{ij} \equiv -\frac{1}{2k_V} \left( D_i \mathcal{V}_j + D_j \mathcal{V}_i \right)~.
\ee
The vectorial harmonics satisfy:
\be
\left( \Delta_{S^2} + k_V^2 \right) \mathcal{V}_{i} = 0~, \quad D_j \mathcal{V}^j = 0~.
\ee
The eigenvalues are given by $k_V^2 = l(l + 1)-1$ with $l = 1,2,\ldots$ being the angular momentum on the $S^2$. Thus, they constitute a single degree of freedom.

Upon defining a master variable $\Phi(r,t) \equiv r^{-1} \Omega(r,t)$ and $(f^a+r D^a H_T /k_V) \equiv r^{-1}\epsilon^{ab}D_b\Omega~$, it is found in \cite{Kodama:2003kk} that the equation satisfied by the master field $\Phi$ is given by:
\be
\square_{g^{(2)}} \Phi - \frac{V_V}{(1-r^2)} \Phi = 0  \implies -(1-r^2)\frac{d}{dr} \left( (1-r^2)\frac{d \Phi}{dr} \right) + V_V \Phi = \omega^2 \Phi~,
\ee
with effective potential:
\be
V_V = \frac{(1-r^2)}{r^2} \left( k_V^2 + 1 \right)~.
\ee
The box operator is the Laplacian corresponding to the two-dimensional metric $g_{ab}$ with $a,b \in \{ t, r \}$. We have further assumed an oscillatory time behavior $\Phi(r,t) = e^{-i\omega t}\varphi(r)$ for the modes. For convenience, the subscript labels $\omega$ and $l$ for the fields have been suppressed.

\subsection{Solution near the origin}

The solution that is well behaved near $r = 0$ is:
\be
\varphi^S(r) = r^{l + 1} (1-r^2)^{-i\omega/2} F\left(a,b;c;r^2 \right)~,
\ee
with
\be
a = \frac{1}{2}\left(1+l - i\omega \right)~, \quad b = \frac{1}{2}\left( 2 + l - i \omega \right)~, \quad c = \frac{3}{2} + l~.
\ee
Clearly, there are a set of northern modes defined in the northern patch which are equivalent to the above, except that $t$ runs backwards.

\subsubsection*{Near the cosmological horizon}

Using hypergeometric identities, we can express the above in a way that makes manifest its behavior near the cosmological horizon $r^2 = 1$. Once again, we exploit equation (\ref{hypident}).
Notice that in this case, $c-a-b = i \omega$ and thus we find a linear combination of ingoing and outgoing modes for $\varphi(r)$ near the cosmological horizon. We use the Kruskal coordinates (\ref{kruskS})
in the southern diamond, with $U>0$ and $V<0$. Near the horizon where $r \to 1$ (and $UV\to 0$):
\be
\varphi^S(r)e^{-i\omega t} \sim \alpha_{\omega l}(-V)^{i\omega} + \alpha^*_{\omega l} U^{-i\omega}~.
\ee
The coefficients $\alpha_{\omega l}$ are given by:
\be
\alpha_{\omega l} \equiv \frac{\Gamma(3/2+l)\Gamma(-i\omega)2^{i\omega}}{\Gamma\left(\frac{1}{2}(1+l-i\omega)\right)\Gamma\left(\frac{1}{2}(2+l-i\omega)\right)}~.
\ee

\subsection{Solution near $\mathcal{I}^+$}

We can also build solutions which are smooth in the region $r \in [1,\infty]$ containing $\mathcal{I}^+$. We find two linearly independent solutions:
\begin{eqnarray}
\varphi^{out-} &=& (r^2 - 1)^{- i \omega /2} r^{i \omega} F\left( \frac{1}{2}(1+l-i\omega ),\frac{1}{2}(-l-i\omega);\frac{1}{2};\frac{1}{r^2} \right)~,\\
\varphi^{out+} &=& (r^2 - 1)^{- i \omega /2} r^{-1+i\omega} F\left( \frac{1}{2}(1-l-i\omega),\frac{1}{2}( 2+l-i \omega );\frac{3}{2};\frac{1}{r^2} \right)~.
\end{eqnarray}
Near $\mathcal{I}^+$ the solutions behave like $\varphi^{out-} \sim 1$ and $\varphi^{out+} \sim 1/r + \mathcal{O}(1/r^{3})$ which implies $\Omega \sim r$ and $\Omega \sim 1 + \mathcal{O}(1/r^{2})$. This in turn implies that the falloffs of the graviton itself, i.e. $r^2 H_T(r,t)$, are given by $\sim r^2$ and $\sim r^{-1}$. Thus, as expected, there is a slow falling and fast falling mode in accordance with the Starobinskii expansion \cite{Starobinsky:1982mr}.

\subsubsection*{Near the cosmological horizon}

Once again, using the same hypergeometric identity (\ref{hypident}), we can expand our solutions near the cosmological horizon to find a linear combination of ingoing and outgoing modes. Near the cosmological horizon $UV \to 0$, in the Kruskal coordinates (\ref{kruskout}) with $U>0$ and $V>0$ we find:
\begin{eqnarray}
\varphi^{out-}(r)e^{-i\omega t} &\sim& \beta_{\omega l}V^{i\omega} + \beta^*_{\omega l} U^{-i\omega }~,\\
\varphi^{out+}(r)e^{-i\omega t} &\sim& \gamma_{\omega l}V^{i\omega} + \gamma^*_{\omega l} U^{-i\omega }~.
\end{eqnarray}
The coefficients $\beta_{\omega l}$ and $\gamma_{\omega l}$ are given by:
\begin{eqnarray}
\beta_{\omega l} &\equiv& \frac{\Gamma(1/2)\Gamma(-i\omega)2^{i\omega}}{\Gamma\left(\frac{1}{2}(1+l-i\omega)\right)\Gamma\left(\frac{1}{2}(-l-i\omega)\right)}~, \\
\gamma_{\omega l} &\equiv& \frac{\Gamma(3/2)\Gamma(-i\omega)2^{i\omega}}{\Gamma\left( \frac{1}{2}(2+l-i\omega)\Gamma\left( \frac{1}{2}(1-l-i\omega) \right) \right)}~.
\end{eqnarray}

\subsection{Matching the flux}
The Bogoliubov transformation between the northern and southern modes and the $out\pm$ modes near $\mathcal{I}^+$ can now be obtained by matching the flux across the future horizons of the two static patches. We find:
\be
\left( \begin{array}{ccc}
\Phi^S_{\omega l} \\
\Phi^N_{\omega l}  \end{array} \right) = \bold{B}_{\omega l}
\left( \begin{array}{ccc}
\Phi^{out+}_{\omega l} \\
\Phi^{out-}_{\omega l}  \end{array} \right)~.
\ee
The matrix $\bold{B}_{\omega l}$ is given by:
\be
\bold{B}_{\omega l} = \frac{1}{ (\beta_{\omega l}^* \gamma_{\omega l} - \gamma_{\omega l}^* \beta_{\omega l})} \left( \begin{array}{ccc}
B_{11} & B_{12} \\
B_{21} & B_{22}  \end{array} \right)~
= -2 i \omega \left( \begin{array}{ccc}
B_{11} & B_{12} \\
B_{21} & B_{22}  \end{array} \right)~,
\ee
with:
\begin{eqnarray}
B_{11} &=& -\alpha_{\omega l}^* \beta_{\omega l}~,\\
B_{12} &=& \alpha_{\omega l}^* \gamma_{\omega l} ~, \\
B_{21} &=& \alpha_{\omega l} \beta^*_{\omega l} ~,\\
B_{22} &=&  - \alpha_{\omega l} \gamma^*_{\omega l}~.
\end{eqnarray}

\subsection{Demonic interference for gravitons }

As in the case of the scalar fields, we can tune the demon modes from the northern patch to cancel the non-normalizable graviton modes coming from the southern observatory.
In particular, suppose the southern observer sends a single southern mode $\varphi^S_{\omega l}$, then its non-normalizable component is $\sim B_{12} \Phi^{out-}$. The northern demon will send in a mode with coefficient $-B_{12}/B_{22} =\left(-1\right)^l$ to cancel out the non-normalizable piece. The resultant mode will only contain the $\Phi^{out+}$ (normalizable) mode whose coefficient is given by:
\be
B_{11} - \frac{B_{12}}{B_{22}}B_{21} =
  \frac{i \Gamma\left(\frac{3}{2}+l\right) \Gamma\left(\frac{1}{2} (1-l+i \omega )\right)}{\omega \sqrt{\pi } \Gamma\left(\frac{1}{2} (1+l+i \omega )\right)}
=B_{11} \left(1- \frac{B_{12} B_{21}}{B_{22} B_{11}}\right)= 2 B_{11}
~.
\ee since ${B_{12}~ B_{21}}/\left({B_{22} ~B_{11}}\right) =-1+ \frac{2 \sin{\left(l \pi\right)}}{\sin{\left(l \pi\right)}+ i \sinh{\left( \pi \omega\right)}}=-1$.
The full mode with no growing behavior near $\ip$ is given by
\be\label{fullgraviton}
\phi^S +\left(-1\right)^l \phi^N
\ee
We do not fully understand why this result is so much simpler than that for the light scalar in dS$_3$ studied in the previous section. The modes (\ref{fullgraviton}) are eigenmodes of the dS anitipodal map and therfore must decay at $\mathcal{I}^-$ as well as
$\mathcal{I}^+$. In the appendix, we compute the linearized graviton in  global coordinates and  verify this is indeed the case.

For completeness, we mention here that our result remains the same in the case of the scalar harmonic perturbations, since the effective equation $V_S$ governing the scalar master function $\Phi_S$ \cite{Kodama:2003kk} is equivalent to $V_V$ (with $l = 0,1,2,\ldots$ in the scalar case).\footnote{It would be interesting to understand whether such a demonic interference can elucidate the boundary conditions imposed at future infinity for the de Sitter-like spacetimes studied in \cite{Anninos:2009yc,Anninos:2010gh,Anninos:2011vd}.}

\section{Analytic continuation AdS $\to$ dS }

In this section we discuss the bulk-to-bulk two-point functions $G(x,x')$ consistent with our future boundary conditions.  These are analogs of vacuum correlation functions, but since we have not explored herein how to define a quantum theory with acausal boundary conditions we can not realize $G(x,x')$ as $\left\langle 0 \left|\phi(x)\phi(x')\right|0\right\rangle$. The allowed modes such as (\ref{good}) are not a complete set on a spacelike slice so we cannot use them to define a state $\left|0\right\rangle$ on a such a  slice. We regard this as a feature rather than a bug since, as we have argued, such a state is unphysical!

Nevertheless a suitable two-point function $G(x,x')$ can be fully determined for $x\neq x'$  (i.e.  up to an $i\varepsilon$ prescription not considered here) from general principles:  the equation of motion for each argument, dS invariance, fast-falling boundary conditions at $\mathcal{I}^+$ and the Hadamard form of the short distance singularity.  dS-invariance implies $G$ can be written purely in terms of the quantity
\be
P(x,x')=\cos {d(x,x') \over \ell}
\ee
where the geodesic distance $d(x,x')$ between $x$ and $x'$ is imaginary for timelike separations.  To be explicit in planar coordinates \be ds_{dS}^2 = \frac{\ell^2}{\eta^2}\left(-d\eta^2 + d{x_1}^2+dx_2^2\right)\ee
one has
\be P(x,x') = \frac{\eta^2 + {\eta'}^2-(x_1-x_1')^2-(x_2-x'_2)^2}{2\eta\eta'}~.\ee For the example of a scalar of mass $\mu$ in dS$_3$, $G$ obeys:
\be \label{Geom}
(1-P^2)\partial^2_P G(P) - 3 P \partial_P G(P) - \mu^2 \ell^2 G(P) = 0~.
\ee
The solutions to the above equation are hypergeometric functions which generally involve both falloffs near $\mathcal{I}^+$. Choosing the solution which is only fast-falling near $\mathcal{I}^+$ gives us:
\be
G(P) = N\left({2 \over 1+P}\right)^{h_+} F\left(h_+,h_+-\half;2h_+-1;{2 \over 1+P} \right)~,
\ee
where $N$ is a normalization factor. This has singularities at both the coincident point limit $P(x,x)=1$ as well as the antipodal point limit $P(x,x_A) = -1$. The singularity for antipodally located points reflects the acausal character of our construction.

In the standard quantum formulation, the scalar Wightman function  in the Euclidean vacuum, which is the only solution to (\ref{Geom}) with no singularities at the antipodal point, is given by \cite{Bousso:2001mw}:
\be
G_E(P) = \frac{\Gamma(h_+)\Gamma(h_-)}{(4\pi)^{3/2}\Gamma(3/2)} F\left(h_+,h_-;\frac{3}{2};\frac{1+P}{2}\right)~.
\ee
We can write our fast-falling  two-point function $G$ in terms of $G_E(P)$ and its antipodal cousin $G_E(-P)$ as:
\be\label{fal}
G(P) =  G_E(P) - e^{-i\delta} G_E(-P) ~,~~~~~~~\delta \equiv \pi(1-\sqrt{1-\mu^2\ell^2})
\ee
provided we take the so-far undetermined normalization factor to be
\be
N = -\frac{i 2^{-2h_+}}{2\pi}~.
\ee
This clearly guarantees that the short distance singularity of $G$ has the canonical Hadamard form.

The standard quantum formulation of a scalar in dS admits a one parameter family of dS-invariant vacua often referred to as $\alpha$-vacua.  With the exception of the Euclidean vacuum, the Wightman function for all of these vacua has antipodal singularities, and the
short-distance singularity does not take the Hadamard form. Our Green function (\ref{fal}) is not the Wightman function in any of the dS $\alpha$-vacua.

This is related to the observation \cite{Bousso:2001mw,  Spradlin:2001nb}
 that, although  the double analytic continuation of AdS Wightman functions are some kind of dS two-point functions, they are not interpretable as Wightman functions in any dS-invariant state. Instead, they are precisely the two-point functions (\ref{fal}). To see this note that the Wightman function
 $G_\x (x,x')$ for a scalar of mass $\mu_\x$ in AdS$_3$ with radius $\ell_\x$ obeys
 \be \label{Gom}
(1-P_\x^2)\partial^2 G_\x(P_\x) - 3 P_\x \partial G(P_\x) + \mu_\x^2 \ell_\x^2 G_\x(P_\x) = 0~.
\ee
with
\be P_\x=\cos {id_\x \over \ell_\x} \ee
 constructed from the geodesic distance $d_\x$ between two points $(x,x')$ in AdS$_3$. Explicitly in Poincare coordinates, \be ds_\x^2 = \frac{\ell_\x^2}{z^2}\left(dz^2 - dt^2 + dy^2\right), \ee we have $P_\x(x,x') = (z^2 + {z'}^2-(t-t')^2 + (y-{y'})^2)/(2 z z')$. Under double analytic continuation $z\to \eta,~~t \to x_1,~~y\to i x_2$ together with $\ell_\x\to i\ell$, we have $ds_\x^2 \to ds_{dS}^2$ and $P_\x\to P$.
Taking
\be P_\x=P,~~~~\ell_\x=i\ell, ~~~~~\mu_\x=\mu, ~~~G_\x=G,
 \ee
then (\ref{Gom}) becomes exactly (\ref{Geom}).  The Hadamard-normalized solution picked out by the standard fast spatial falloff  in AdS is  then \be
G_\x(P_\x) = N\left({2 \over 1+P_\x}\right)^{h_+} F\left(h_+,h_+-\half;2h_+-1;{2 \over 1+P_\x} \right)~,
\ee
 where here $h_+=1+\sqrt{1+\mu^2_\x\ell^2_\x}$. Hence  double analytic continuation
 maps the standard AdS Wightman function to the two-point function (\ref{fal}) consistent with future dS boundary conditions.

 Note that both the dS and AdS two-point function have singularities at $P=-1$ which does not correspond to coincident (or null-separated) points. As discussed above, for the dS case this is an
 acausal singularity for spacelike antipodally separated points. In AdS, $P=-1$ corresponds to two noncoincident, timelike separated points connected by a light ray which is reflected off of the AdS boundary.

\section*{Acknowledgements}

It has been a great pleasure discussing this work with L.~Andersson, A.~Castro, J.~Hartle, S.~Hartnoll, S.~Hawking,  D.~Hofman, J.~Maldacena, A.~Maloney, D.~Marolf, D.~Page and S.~Shenker. This work was supported in part by DOE grant DE-FG02-91ER40654 and the Fundamental Laws Initiative at Harvard.

\appendix

\section{Graviton in the global patch}

We consider the global patch of dS$_4$:
\be
\frac{ds^2}{\ell^2} = -d\tau^2 + \cosh^2\tau d\Omega_3^2~,
\ee
covering the full space. In global conformal coordinates one has:
\be
\frac{ds^2}{\ell^2} = \frac{1}{\cos^2{T}}\left[-dT^2 + d\Omega_3^2 \right]~.
\ee
We take the parametrization of the three-sphere to be
\be
d\Omega_3^2 = \frac{1}{(1-r^2)} dr^2 + r^2 d\Omega_2^2~, \quad r \in [0,1]~.
\ee
and defined $\cos{T} \equiv \left(\cosh{\tau}\right)^{-1}$. Gravitational perturbations about this background has been analysed in \cite{KodamaGrav}. In their notation, $a(T) \equiv \left(\cos{T}\right)^{-1}$.

We focus on the tensor harmonics $Y_{(T)ij}$ of the three-sphere, obeying the transverse-traceless condition. These give rise to 2 independent degrees of freedom. The equation to be solved is given by \cite{KodamaGrav}:
\be
\delta R_{ij}-\frac{3}{\ell^2}\delta g_{ij} = 0~.
\ee
Parametrizing the perturbation as $\delta g_{ij} = 2  a(T)^2  H(T) Y_{(T)ij}$, the linearized Einstein's equation becomes:
\begin{eqnarray}
0& = & H'' + 2  \Omega H' + 2 \Omega' H + 4 \Omega^2 H + \left(k_T^2 + 6\right)H -6 a^2 H \\
\Rightarrow 0&=&H''+ 2\tan{T} H'+\left(2+k_T^2\right) H
\end{eqnarray}
with $\Omega(T) \equiv a'(T)/a(T)$ and $k_T^2 = l\left(l+2\right) -2~,~l=1,2,\ldots$

The two independent solutions are then given by:
\begin{eqnarray}
H^{(0)} &=& F\left(-\half - K,-\half + K;-\frac{1}{2};\frac{1}{\cosh^{2}{\tau}}\right) \\
H^{(3)} &=& \frac{1}{\cosh^{3}{\tau}}F\left(1-K,1+K;\frac{5}{2};\frac{1}{\cosh^{2}{\tau}}\right)~,
\end{eqnarray}
where:
\be
K\equiv \frac{\sqrt{k_T^2+3}}{2}~.
\ee
Recall that $\delta g_{ij} \sim a^2 H(\tau) = \left(\cosh{\tau}\right)^2 H(\tau)$, we note the following Starobinskii fall-offs \cite{Starobinsky:1982mr} near $\ip$:
\begin{eqnarray}
 a(\tau)^2 H^{(0)} &\sim& e^{2\tau} + \ldots~, \\
 a(\tau)^2 H^{(3)} &\sim& e^{-\tau} + \ldots~,
\end{eqnarray}
while at ${\cal I}^-$ a similar behavior is observed:
\begin{eqnarray}
 a(\tau)^2 H^{(0)} &\sim& e^{2\tau} + \ldots~, \\
 a(\tau)^2 H^{(3)} &\sim& e^{-\tau} + \ldots~
\end{eqnarray}
The first mode is growing {\it both}  at $\ip$ and ${\cal I}^-$ and the second mode is decaying at {\it  both} boundaries. This is due to the fact that these solutions only depend on $\tau$ through $\cosh{\tau}$ which is invariant under time reversal $\tau \rightarrow -\tau$. Hence the decay behavior near one boundary is the same as the behavior near the other boundary.


\begin{thebibliography}{1}
\bibitem{Gibbons:1977mu}
  G.~W.~Gibbons, S.~W.~Hawking,
  ``Cosmological Event Horizons, Thermodynamics, and Particle Creation,''
  Phys.\ Rev.\  {\bf D15}, 2738-2751 (1977).


\bibitem{Alishahiha:2004md}
  M.~Alishahiha, A.~Karch, E.~Silverstein, D.~Tong,
  ``The dS/dS correspondence,''
  AIP Conf.\ Proc.\  {\bf 743}, 393-409 (2005).
  [hep-th/0407125].

\bibitem{Banks:2003cg}
  T.~Banks,
  ``Some thoughts on the quantum theory of de sitter space,''
  [astro-ph/0305037].

\bibitem{Banks:2010tk}
  T.~Banks,
  ``Pedagogical notes on black holes, de Sitter space, and bifurcated horizons,''
  [arXiv:1007.4003 [hep-th]].


\bibitem{Goheer:2002vf}
  N.~Goheer, M.~Kleban, L.~Susskind,
  ``The Trouble with de Sitter space,''
  JHEP {\bf 0307}, 056 (2003).
  [hep-th/0212209].

\bibitem{Castro:2011xb}
  A.~Castro, N.~Lashkari, A.~Maloney,
  ``A de Sitter Farey Tail,''
  [arXiv:1103.4620 [hep-th]].


\bibitem{Banks:2002wr}
  T.~Banks, W.~Fischler, S.~Paban,
  ``Recurrent nightmares? Measurement theory in de Sitter space,''
  JHEP {\bf 0212}, 062 (2002).
  [hep-th/0210160].

\bibitem{Witten:2001kn}
  E.~Witten,
  ``Quantum gravity in de Sitter space,''
  [hep-th/0106109].

\bibitem{Anninos:2010zf}
  D.~Anninos, G.~S.~Ng, A.~Strominger,
  ``Asymptotic Symmetries and Charges in De Sitter Space,''
  [arXiv:1009.4730 [gr-qc]].


\bibitem{BM62}
  H.~Bondi, M.~G.~J.~van der Burg and A.~W.~K.~Metzner,
   ``Gravitational waves in general relativity. 7. Waves from axisymmetric
  isolated systems,''
  Proc.\ Roy.\ Soc.\ Lond.\  A {\bf 269}, 21 (1962).

\bibitem{S62a}
  R.~K.~Sachs,
   ``Gravitational waves in general relativity. 8. Waves in asymptotically flat
  space-times,''
  Proc.\ Roy.\ Soc.\ Lond.\  A {\bf 270}, 103 (1962).

\bibitem{TW66}
L.~A.~Tamburino and J.~H.~Winicour,
  ``Gravitational Fields in Finite and Conformal Bondi Frames,''
 Phys. Rev. 150, 10391053 (1966)

\bibitem{Maldacena:2002vr}
  J.~M.~Maldacena,
  ``Non-Gaussian features of primordial fluctuations in single field inflationary models,''
  JHEP {\bf 0305}, 013 (2003).
  [astro-ph/0210603].

\bibitem{Strominger:2001pn}
  A.~Strominger,
  ``The dS / CFT correspondence,''
  JHEP {\bf 0110}, 034 (2001).
  [hep-th/0106113].

\bibitem{Strominger:2001gp}
  A.~Strominger,
  ``Inflation and the dS / CFT correspondence,''
  JHEP {\bf 0111}, 049 (2001).
  [hep-th/0110087].

\bibitem{Maldacena:2011mk}
  J.~Maldacena,
  ``Einstein Gravity from Conformal Gravity,''
  [arXiv:1105.5632 [hep-th]].


\bibitem{es} E. Schrodinger, ``Expanding Universes,'' Cambridge University Press, 1956.

\bibitem{Parikh:2002py}
  M.~K.~Parikh, I.~Savonije, E.~P.~Verlinde,
  ``Elliptic de Sitter space: dS/$Z_2$,''
  Phys.\ Rev.\  {\bf D67}, 064005 (2003).
  [hep-th/0209120].

\bibitem{Sanchez:1987sa}
  N.~Sanchez,
  ``Quantum Field Theory and the `Elliptic Interpretation' of de Sitter Space-Time,''
  Nucl.\ Phys.\ {\bf B294}, 1111 (1987).


\bibitem{Horowitz:2003he}
  G.~T.~Horowitz and J.~M.~Maldacena,
  ``The Black hole final state,''
  JHEP {\bf 0402}, 008 (2004)
  [arXiv:hep-th/0310281].

 \bibitem{Susskind:1993if}
  L.~Susskind, L.~Thorlacius, J.~Uglum,
  ``The Stretched horizon and black hole complementarity,''
  Phys.\ Rev.\  {\bf D48}, 3743-3761 (1993).
  [hep-th/9306069].

\bibitem{Bousso:2001mw}
  R.~Bousso, A.~Maloney, A.~Strominger,
  ``Conformal vacua and entropy in de Sitter space,''
  Phys.\ Rev.\  {\bf D65}, 104039 (2002).
  [hep-th/0112218].


\bibitem{Kodama:2003kk}
  H.~Kodama, A.~Ishibashi,
  ``Master equations for perturbations of generalized static black holes with charge in higher dimensions,''
  Prog.\ Theor.\ Phys.\  {\bf 111}, 29-73 (2004).
  [hep-th/0308128].



\bibitem{Starobinsky:1982mr}
  A.~A.~Starobinskii,
  ``Isotropization of arbitrary cosmological expansion given an effective cosmological constant,''
  JETP Lett.\  {\bf 37}, 66-69 (1983).


\bibitem{Anninos:2009yc}
  D.~Anninos, T.~Hartman,
  ``Holography at an Extremal De Sitter Horizon,''
  JHEP {\bf 1003}, 096 (2010).
  [arXiv:0910.4587 [hep-th]].

\bibitem{Anninos:2010gh}
  D.~Anninos, T.~Anous,
  ``A de Sitter Hoedown,''
  JHEP {\bf 1008}, 131 (2010).
  [arXiv:1002.1717 [hep-th]].

\bibitem{Anninos:2011vd}
  D.~Anninos, S.~de Buyl, S.~Detournay,
  ``Holography For a De Sitter-Esque Geometry,''
  JHEP {\bf 1105}, 003 (2011).
  [arXiv:1102.3178 [hep-th]].

\bibitem{Spradlin:2001nb}
  M.~Spradlin and A.~Volovich,
  ``Vacuum states and the S matrix in dS / CFT,''
  Phys.\ Rev.\  D {\bf 65}, 104037 (2002)
  [arXiv:hep-th/0112223].


 \bibitem{KodamaGrav}
  H.~Kodama, M.~Sasaki,
  ``Cosmological Perturbation Theory,''
  Prog. Theor. Phys. Suppl. No. 78 (1984), Appendix D.

\end{thebibliography}
\end{document}